# V-Doped Niobate Nanosheets for Enhanced Photocatalytic Activity

M. Tuğrul Avcu[†], Uğur Ünal*[‡§#]

† Materials Science and Engineering Department, Koç University, 34450 Istanbul, Turkey

‡ Chemistry Department, Koç University, 34450 Istanbul, Turkey

§ Koç University Surface Science and Technology Centre (KUYTAM), 34450 Istanbul, Turkey

# Koç University Hydrogen Technologies Center (KUHyTech), 34450 Istanbul, Turkey

**ABSTRACT:** V-doped $[Ca_2Nb_{3-x}V_xO_{10}]^-$ (x= 0, 0.15, 0.3, 0.6, 0.75) nanosheets were produced by solid state reaction followed by protonation and exfoliation by tetrabutylammonium ions. 2D nanosheets have improved photocatalytic activity due to their small thickness that reduces electron-hole recombination, larger surface area that increases photocatalytic sites, and decreased band gap from added V. V-doping has caused the wide band gap of 3.54 eV of undoped nanosheets to decrease to 2.60-2.88 eV range. 10 vol% methanol was used as a hole scavenger in the hydrogen evolution reactions, and 3 wt% Pt was deposited on nanosheets as a co-catalyst. Under a full-spectrum Xe light, $[Ca_2Nb_{2.7}V_{0.3}O_{10}]^-$ has produced 4.7 times the $H_2$ yield as undoped nanosheets with a 11.3 mmol/g/h production rate, $[Ca_2Nb_{2.7}V_{0.3}O_{10}]^-$ with Pt co-catalyst has produced 2.9 times the $H_2$ yield as undoped Pt-loaded nanosheets.

## INTRODUCTION

The photocatalytic water splitting into hydrogen and oxygen using a $TiO_2$ electrode has been a landmark in semiconductor photocatalysis.[1] This triggered intensive research on oxide-based electrocatalysts, especially perovskite-structured $SrTiO_3$,[2–5] whose favorable band edge positions and chemical stability made it an effective candidate for water splitting. Following studies expanded the material scope into a wide range of metals, oxides and oxide derivatives, establishing clear relationships between their crystal structure and electronic properties.[6–11] Titanate perovskites were later shown to function as freestanding photocatalysts without the need of an electrode system,[12] branching further studies of photocatalysis.[13–17]

Ruddlesden-Popper ($A'_2A_{n-1}B_nO_{3n+1}$) and Dion-Jacobson ($A'A_{n-1}B_nO_{3n+1}$) phase perovskite oxides[18,19] (LPOs) were important as their interlayer chemistry enables ion exchange of their A' cation with protons, emphasizing their solid-acid character, and chemical tunability. Solid-acid character of these proton-exchanged LPOs enabled the studies investigating their acidic properties, and reactions with organic bases.[16,20–26] Intercalation of bulky organic cations and deprotonation under basic conditions increases interlayer repulsion, inducing swelling and eventually exfoliation into two-dimensional nanosheets.[27,28] These 2D structures offer enhanced charge transport and abundant photocatalytic sites, making them highly attractive platforms for photocatalysis.[29–31]

Performance can be further improved through co-catalyst loading and B-site substitution to directly modify the electronic structure.[6,12–14,25,32,33] One of the biggest challenges that these LPOs pose is their wide band gap, and combination of transition metals with p-block elements, such as nitridation, enables engineering their band structures.[34–37] However, oxynitride structures have low photocatalytic activities under visible light irradiation and their chemical stability needs to be evaluated further.[36–38] Since nitridation potentially compromises chemical stability while causing low photocatalytic performance in the visible range, alternative strategies must be explored.

Vanadium and its oxide derivatives have attracted interest across a broad range of catalytic applications. $V_2O_5$ has emerged as a visible light photocatalyst for water splitting, and $V_2O_5$-based composites have shown promise as electrocatalysts for both oxygen and hydrogen evolution reactions.[39–44] V-doping has also proven to be an effective band gap engineering strategy in various oxide structures, shifting optical absorption into the visible region due to its low band gap of ~2.20 eV and enhancing photocatalytic activity.[44–49] Therefore, this study investigates V substitution at the B-site of the Dion-Jacobson phase perovskite, $KCa_2Nb_3O_{10}$, as a novel strategy for band gap engineering. The effects of V incorporation into the crystal structure, optical absorption behavior, and electronic properties of the V-doped LPO were examined. V-doped nanosheets derived from the parent LPO were prepared, and their photocatalytic performance were evaluated in relation to the modified electronic structure and light response.

## EXPERIMENTAL SECTION

**Nanosheet Preparation.** The LPO $KCa_2Nb_{3-x}V_xO_{10}$ (x = 0, 0.15, 0.3, 0.6, 0.75; abbreviated as KN3, KV5, KV10, KV20, KV25) powders were prepared by solid-state reaction of $K_2CO_3$ (99.0%, Alfa Aesar) (65% mol excess), $CaCO_3$ (98.5-100.5%, Merck), $Nb_2O_5$ (99.9%, Sigma-Aldrich), and V2O5 (99.6%, Sigma-Aldrich). Excess $K_2CO_3$ was used to compensate for the volatilization of $K_2CO_3$, with a melting point of 890 °C, and various potassium vanadate compounds such as $KVO_3$, $K_3V_5O_{14}$, $K_2V_2O_7$ with a melting point range of 410-520 °C, that could form as intermediates during the calcination. Calcination was done at 1200 °C for 10 h with a heating rate of 10 °C $min^{-1}$, and natural cooling. Proton exchanged $HCa_2Nb_{3-x}V_xO_{10}$ perovskites (x = 0, 0.15, 0.3, 0.6, 0.75; abbreviated as HN3, HV5, HV10, HV20, HV25) were obtained by stirring the parent oxides in 5 M $HNO_3$, with a 1:250 stoichiometric ratio of $K^+$:$H_3O^+$, for 3 days. After the proton exchange process, the powders were washed several times and dried in room conditions or 80 °C. Nanosheet suspensions were obtained by stirring the proton exchanged forms in 10 mM tetrabutylammonium hydroxide ($TBA^+OH^-$) (40 wt%, Sigma-Aldrich) for 7 days, with a 1:1 stoichiometric ratio of $HCa_2Nb_{3-x}V_xO_{10}$:$TBA^+$. After 7 days, the mixtures were left undisturbed for 1 day to allow for the unexfoliated powders to precipitate, and the supernatant was collected as the nanosheet suspension. Nanosheets were restacked by adding 1 M HCl until precipitation, precipitated mixtures were washed several times, and their supernatants were discarded. Wet precipitates were freeze-dried to obtain $[Ca_2Nb_{3-x}V_xO_{10}]^-$ nanosheet powders (x = 0, 0.15, 0.3, 0.6, 0.75; abbreviated as N3, V5, V10, V20, V25) with concentrations ranging from 2.2 to 4.3 mg/mL.

For photoelectrochemical (PEC) characterizations, nanosheet suspensions were dip coated onto 1 cm×1 cm FTO substrates using a 2.5 g/L polyethyleneimine (branched, Mw ~25000, Mn ~10 000, Sigma-Aldrich) solution. Prior to coating, FTO substrates were cleaned by 254 nm UV light and ozone. The substrates were dipped in PEI for 15 min, deionized water for 5 min, nanosheet suspension for 15 min, and deionized water for 5 min, for a total of 10 layers. Obtained films were then dried at 80 °C.

**Characterization and Equipment.** Crystal structures of all forms were analyzed by X-ray powder diffraction (XRD) using a Rigaku MiniFlex 600 diffractometer with Cu Kα radiation. Raman analyses were done using 633 nm excitation laser with a Renishaw, Invia Reflex. Morphological analysis was done by field emission scanning electron microscopy (FESEM; Zeiss, Ultra Plus). UV-Vis absorption spectra were obtained by diffuse reflectance spectroscopy using a Shimadzu UV-3600 Plus. X-ray photoelectron spectroscopy (XPS) was conducted on a Thermo Scientific K-Alpha spectrometer using monochromatic Al Kα excitation. Obtained XPS data were analyzed using the Thermo Avantage v5.973 software. Photocurrent responses of nanosheet-coated FTO substrates were measured by chronoamperometry (CA) using a Princeton Applied Research Versa STAT MC potentiostat in a three-electrode electrochemical cell. Nanosheet films were used as working electrodes, a spiral Pt wire was used as the counter electrode, and saturated calomel electrode (SCE) was used as a reference electrode. 0.5 M $Na_2SO_4$ (pH=7) was used as an electrolyte and was purged with Ar flow before the experiments. 300 W Xe lamp (Newport, 6259) was used in photocatalytic hydrogen evolution reactions (HER) and PEC experiments. Hydrogen amounts in HER experiments were measured with a Shimadzu 2014 gas chromatograph (GC).

**Photocatalytic Experiments.** Photocatalytic HER was done in a 10 mL quartz cell using 10 vol% methanol-water solution. 15 mg of bulk or proton exchanged powder was added in 5 mL of solution or 2.5 mg of nanosheet powder was added in 2.5 mL of solution. To load 3 wt% Pt co-catalyst onto nanosheets, $PtCl_4$ (57.5% Pt, Merck) solution was added into the solution and irradiated with the Xe lamp for 30 min. The mixtures were sonicated for 30 min and purged with 50 sccm Ar flow for 30 min before the experiments. Samples were taken in the first 30 min and every hour for 6h.

## RESULTS AND DISCUSSION

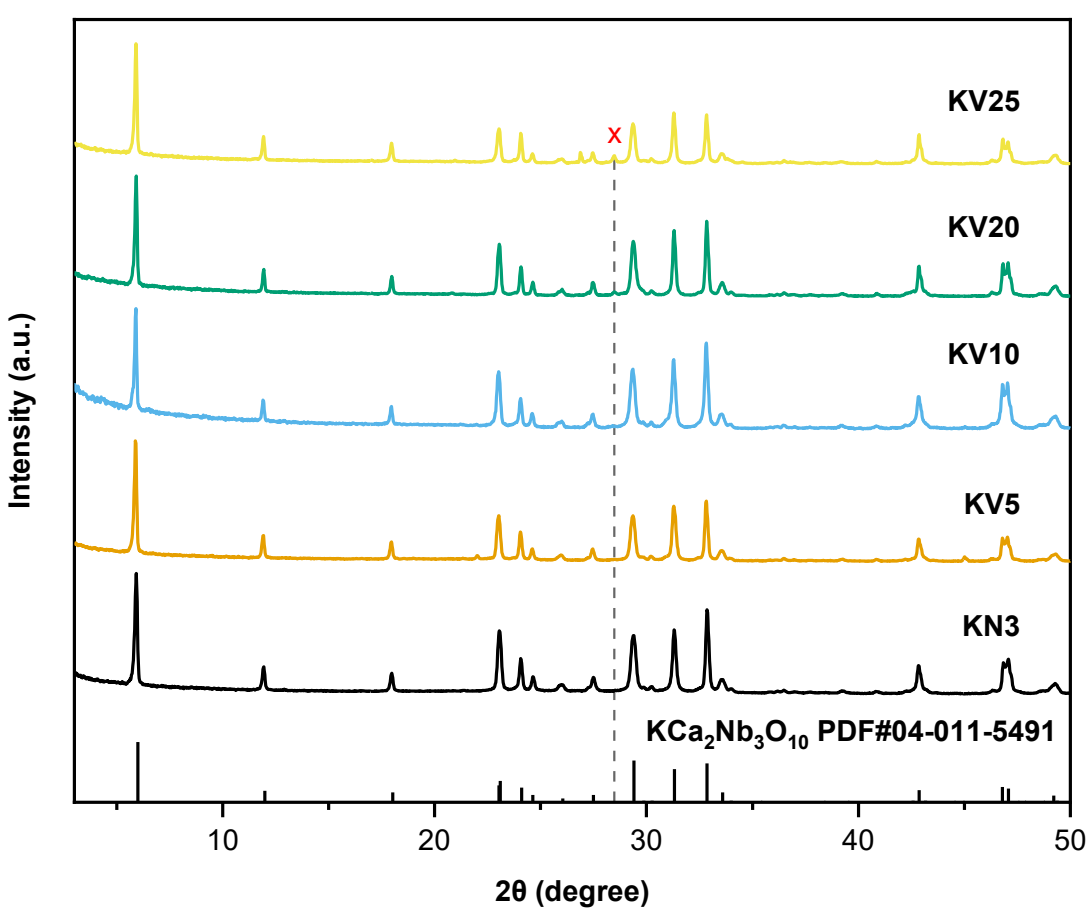

Figure 1. XRD diffractograms of the parent $KCa_2Nb_{3-x}V_xO_{10}$ compositions. Secondary phase of (x) $KVO_3$ is indicated.

Figure 1 shows XRD patterns of different $KCa_2Nb_{3-x}V_xO_{10}$ compositions with orthorhombic symmetry and Cmcm space group.[50] The peak at 29° due to the formation of $KVO_3$ secondary phase starts to appear after 20% V-doping at the B-site (x = 0.6), mostly due to excess K amount, and becomes more pronounced by increased doping amount. The (020) peak corresponding to layer thickness around 6° has values of 5.92, 5.88, 5.90, 5.92, and 5.90°, respectively with increasing doping amount. As $V^{5+}$ and $Nb^{5+}$ with 6 coordination number have ionic radii of 68 and 78 pm respectively, a decrease in layer thickness and thus, an increase in the (020) peak was meant to be observed like in oxynitrides but unexpectedly, that did not happen. By adding a smaller ion at the B-site, system entropy increases and this may lead to a distortion of the octahedra in the lattice, causing an opposite effect to what was expected.

XRD patterns of parent, dried proton exchanged, and nanosheet forms of all V-doped compositions are given in Figure S1. After proton exchange, the space group changes to P4/mmm and more than one layer of hydronium ions intercalates into the interlayer space.[51] Drying the proton exchanged powders in room conditions helps maintain the extra layer of $H_3O^+$, causing an increase in photocatalytic performance due to the readily available hydrogens in the interlayer. Drying the proton exchanged forms at 80 °C causes the extra layer to evaporate and space group is claimed to change into P4/mbm,[52] or not change at all.[51] However, lack of hydronium ions in the interlayer leads to lower $H_2$ production during HER. In this study, for the sake of the consistency of instrumental analyses, the dried form of the proton exchanged powders were used for characterizations. For the dried proton exchanged powders, (020) peak indicating interlayer distance, has a higher theoretical 2θ value of 6.124° due to the exchange of $K^+$ (r = 133 pm) ions with $H_3O^+$ (r = 100 pm). Patterns of the nanosheet forms show that only (020), (002), (112), and (004) peaks remain after restacking indicating a disordered system. (020) peaks decrease due to presence of $TBA^+$ ions in the interlayer space.

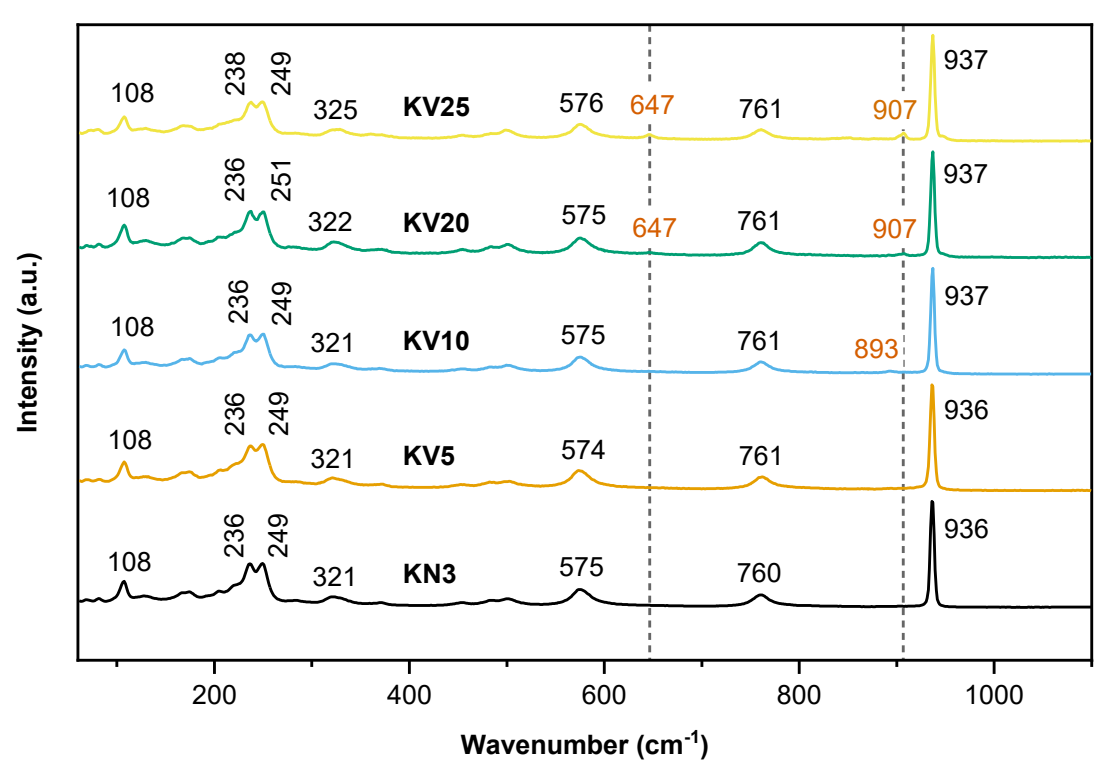

Figure 2. Raman spectra of $KCa_2Nb_{3-x}V_xO_{10}$ compositions.

Raman spectra reveal the stretching mode of terminal M-O bonds at 936 $cm^{-1}$, weakly distorted central $MO_6$ octahedra at 760 $cm^{-1}$, highly distorted central $MO_6$ octahedra at 575 $cm^{-1}$, stretching mode of M-O at 320 $cm^{-1}$, bending mode of M-O-M bonds at 240 $cm^{-1}$, and low-frequency lattice modes associated with the stacking of perovskite layers below 200 $cm^{-1}$.[53] As presented in Figure 2, the peak at 321 $cm^{-1}$ shifts to 322 and 325 $cm^{-1}$ with increasing V content while the other peaks remain unchanged. After 20% V-doping, peaks at 647 and 907 $cm^{-1}$ emerge, confirming V presence in both terminal and central octahedra. However, a weak intensity peak at 893 $cm^{-1}$ instead of 907 $cm^{-1}$ is observed for the KV10 composition.

Raman spectra of different $HCa_2Nb_{3-x}V_xO_{10}$ compositions (Figure S2) show that increased V-doping causes a slight increase in the 317 $cm^{-1}$ peak and a slight decrease in the 772 $cm^{-1}$ peak. The double peak of the parent compounds at around 240 $cm^{-1}$ shifts to 216 $cm^{-1}$ and becomes a single peak. With proton exchange, the peaks at 937 $cm^{-1}$ decrease in intensity dramatically and shifts to 968 $cm^{-1}$, confirming a change in interlayer chemistry, affecting the terminal M-O bonds. However, the same peak of undoped HN3 is even smaller and shifted to a higher wavenumber compared to the V-doped LPOs. Also, the 575 $cm^{-1}$ peak of HN3 shifts to higher wavenumbers with V-doping.

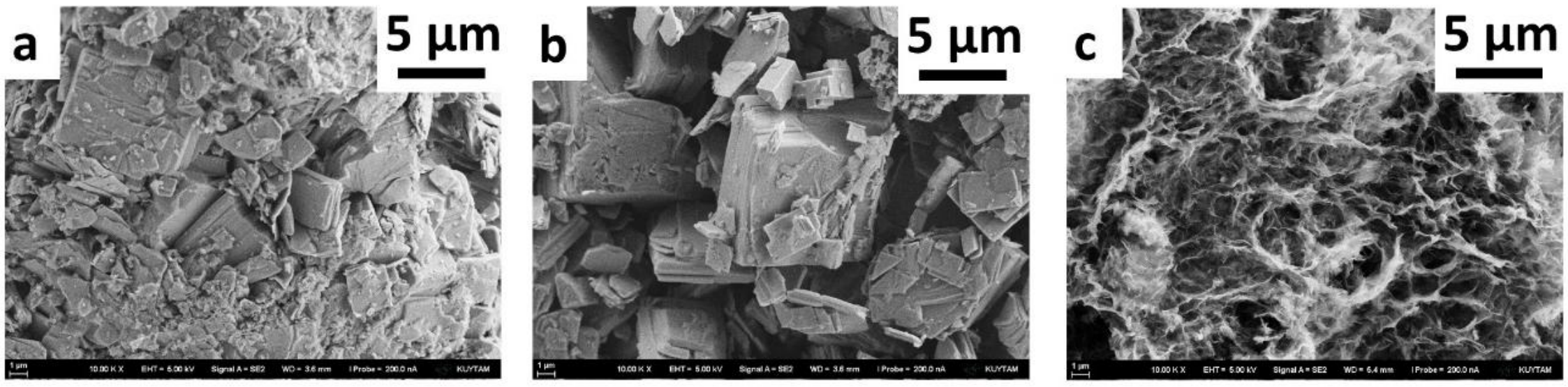

Figure 3. SEM images of (a) KV10, (b) HV10, and (c) V10.

SEM images of parent, proton exchanged, and nanosheet forms of $KCa_2Nb_{3-x}V_xO_{10}$ phases are given in Figure 3 andFigure S3. V-doped materials have larger particle size compared to the undoped compound, however, a change in surface morphology is not observed due to V-doping or proton exchange. Crumpled surfaces with large surface area, and individual layers of atomic thickness are observed after exfoliation due to the random restacking of nanosheets.

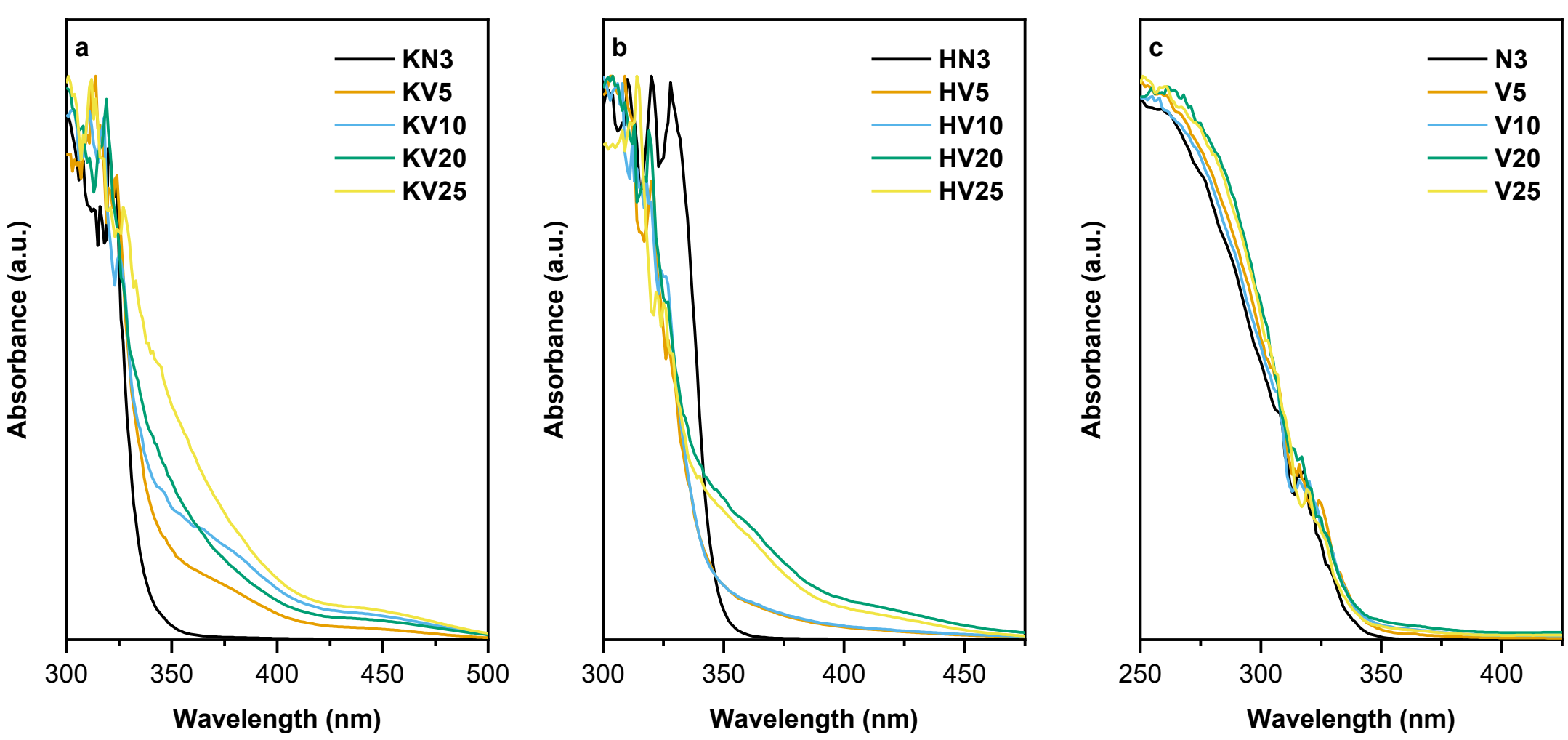

Figure 4. UV-Vis absorbance spectra of (a) parent, (b) proton exchanged, and (c) nanosheet forms of $KCa_2Nb_{3-x}V_xO_{10}$ compositions.

Figure 4 shows the absorption spectra obtained by using the Kubelka-Munk function on reflectance spectra. In parent and proton exchanged forms, V-doping and absorbance limits are seen to be loosely related. However, in the nanosheet forms, the absorbance limit does not change significantly with V-doping, excluding the 350-375 nm region.

Linear fits were applied to the Tauc plots of all compositions for the band gap calculations (Figure S4). V-doping introduces a new conduction band position of V3p which is confirmed from these Tauc plots. Two linear regions correspond to the Nb-O and V-O bands. Band gaps for Nb-O bonds are in the 3.29-3.60 eV range, and for V-O bonds, band gaps range from 2.60-2.94 eV. In proton exchanged and nanosheet forms, both band values seem to decrease, while a minimum value of 2.60 eV is achieved with the V20 composition.

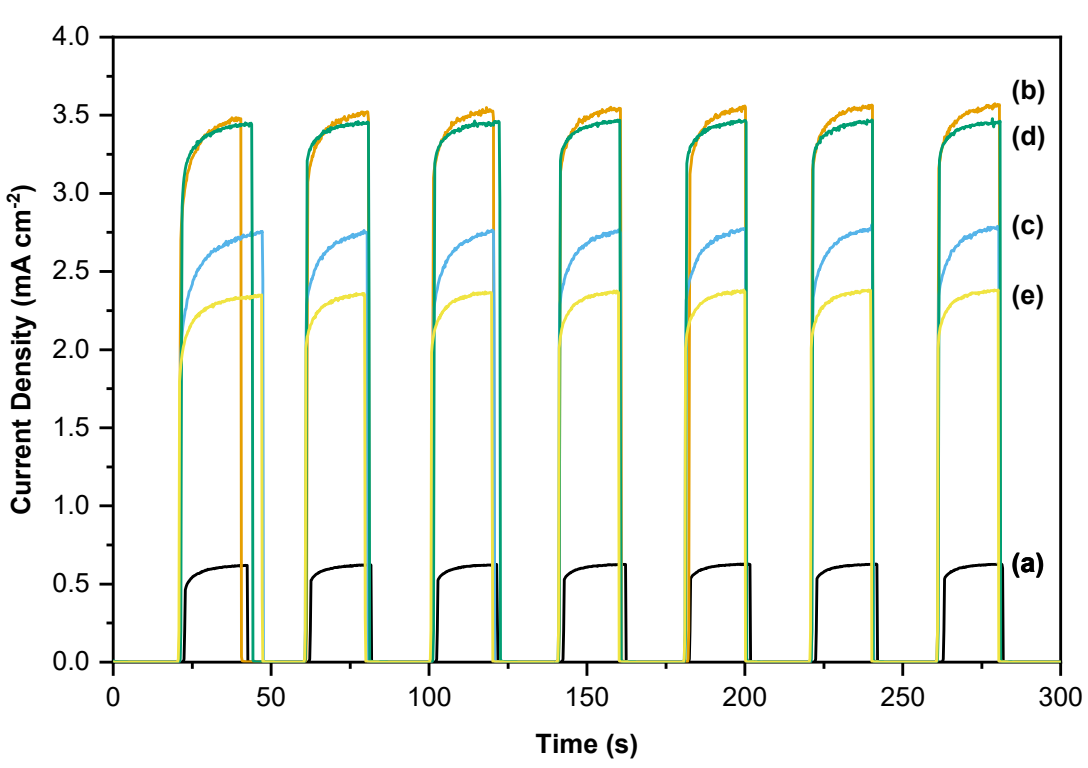

Figure 5. Photocurrent responses measured by CA of (a) N3, (b) V5, (c) V10, (d) V20, and (e) V25 nanosheet films.

Photocurrent responses were measured by chronoamperometry (CA) to evaluate the photoelectrochemical performance and stability during a 300 s experiment with 20 s light on-off cycles (Figure 5). All nanosheet samples exhibited increased photocurrent performance with V-doping compared to the undoped sample due to the decreased band gap of V-doped perovskites. While V-doping increases the photocurrent, a direct correlation between the doping amount and current density is not observed. Current density of N3 is measured to be 629 μA $cm^{-2}$ at its peak, and V-doped samples of V5, V10, V20, and V25 generated 3573, 2800, 3478, and 2382 μA $cm^{-2}$, respectively. All nanosheet films show no photocurrent decay at the beginning of their cycle, suggesting negligible surface recombination due to efficient charge separation and minimal surface trap states. Removing the external potential applied via the light source leads to rapid recombination and photocurrent decay in milliseconds, indicating fast recombination without external bias and further confirming low trap state density of these nanosheets.

Table 1. $H_2$ production rate at the end of the 6 h experiments for all phases of $KCa_2Nb_{3-x}V_xO_{10}$ compositions.

| Perovskite Composition | $H_2$ evolution rate of parent compounds (μmol/g/h) | $H_2$ evolution rate of proton exchanged compounds (μmol/g/h) | $H_2$ evolution rate of nanosheets (mmol/g/h) | $H_2$ evolution rate of Pt-loaded nanosheets (mmol/g/h) |
|---|---|---|---|---|
| $KCa_2Nb_3O_{10}$ | 28.7 | 122 | 2418 | 5522 |
| $KCa_2Nb_{2.85}V_{0.15}O_{10}$ | 1.04 | 707 | 4300 | 12843 |
| $KCa_2Nb_{2.7}V_{0.3}O_{10}$ | 0.796 | 536 | 11336 | 16082 |
| $KCa_2Nb_{2.4}V_{0.6}O_{10}$ | 1.14 | 1192 | 2813 | 4279 |
| $KCa_2Nb_{2.25}V_{0.75}O_{10}$ | 1.44 | 1066 | 2999 | 4167 |

Amount of $H_2$ evolved during the experiment (in Figure S5) and $H_2$ production rates (in

Table 1) are shown for each sample. Due to the inactivity of parent compounds, not every sample taken from the cell gives a detectable amount of $H_2$. However, proton exchanged compounds show 2 to 3 orders of magnitude difference in V-doped materials. While among the parent compounds, undoped perovskite is more productive than its V-doped counterparts, among the proton exchanged materials, V-doped samples produce more $H_2$ than undoped proton exchanged perovskite. Exfoliation into nanosheets seems to provide the undoped nanosheet structure a bigger leap for photocatalytic activity. Undoped nanosheet powder produces 20× $H_2$ compared to its proton exchanged counterpart, while for V-doped samples this amount varies between 2.4-6.1×, except for V10 with a 21× increase. Although undoped nanosheets have such an increase after exfoliation, 10% V-doped nanosheets still produce 4.7× more $H_2$. Furthermore, Pt-loading has a more significant effect on low V-amount V5 nanosheet powder, improving photocatalytic performance by 3.0× compared to around 1.4× increase of other V-doped nanosheets. As a result, Pt-loaded V5 and V10 nanosheet powders are 2.3 and 2.9× as efficient as the Pt-loaded undoped nanosheet powder.

**CONCLUSIONS**

V-doped $[Ca_2Nb_{3-x}V_xO_{10}]^-$ nanosheets were produced for the first time via addition of $V_2O_5$ with solid state reaction, proton exchange, and exfoliation, demonstrating that V-doping is an effective strategy for tuning the optical properties of layered perovskites. Raman spectra of parent compounds revealed that due to V-doping, two new peaks emerge, confirming V presence in both the terminal and central octahedra. V-doping reduced the band gap by the addition of its conduction band to the band structure, shifting the band gap to 2.60-2.88 eV range for nanosheet powders. With V-doping, nanosheet powders have 1.3-4.7× more photocatalytic performance and Pt-loading onto the nanosheet powders affect V-doped samples more. These findings highlight the potential of V-doping for developing stable and efficient visible-light-driven photocatalysts. The results also present a promising approach for reducing the band gap without compromising chemical stability or photocatalytic performance. Future studies focusing on optimization of dopant concentration, co-catalyst loading, long-term stability, and electronic properties may help further our understanding of the effects of V-doping and its role in enhancing photocatalytic hydrogen production.

**ACKNOWLEDGMENTS**

This study was funded by TÜBİTAK (The Scientific and Technological Research Council of Türkiye) under grant no. 123Z415.

**SUPPORTING INFORMATION**

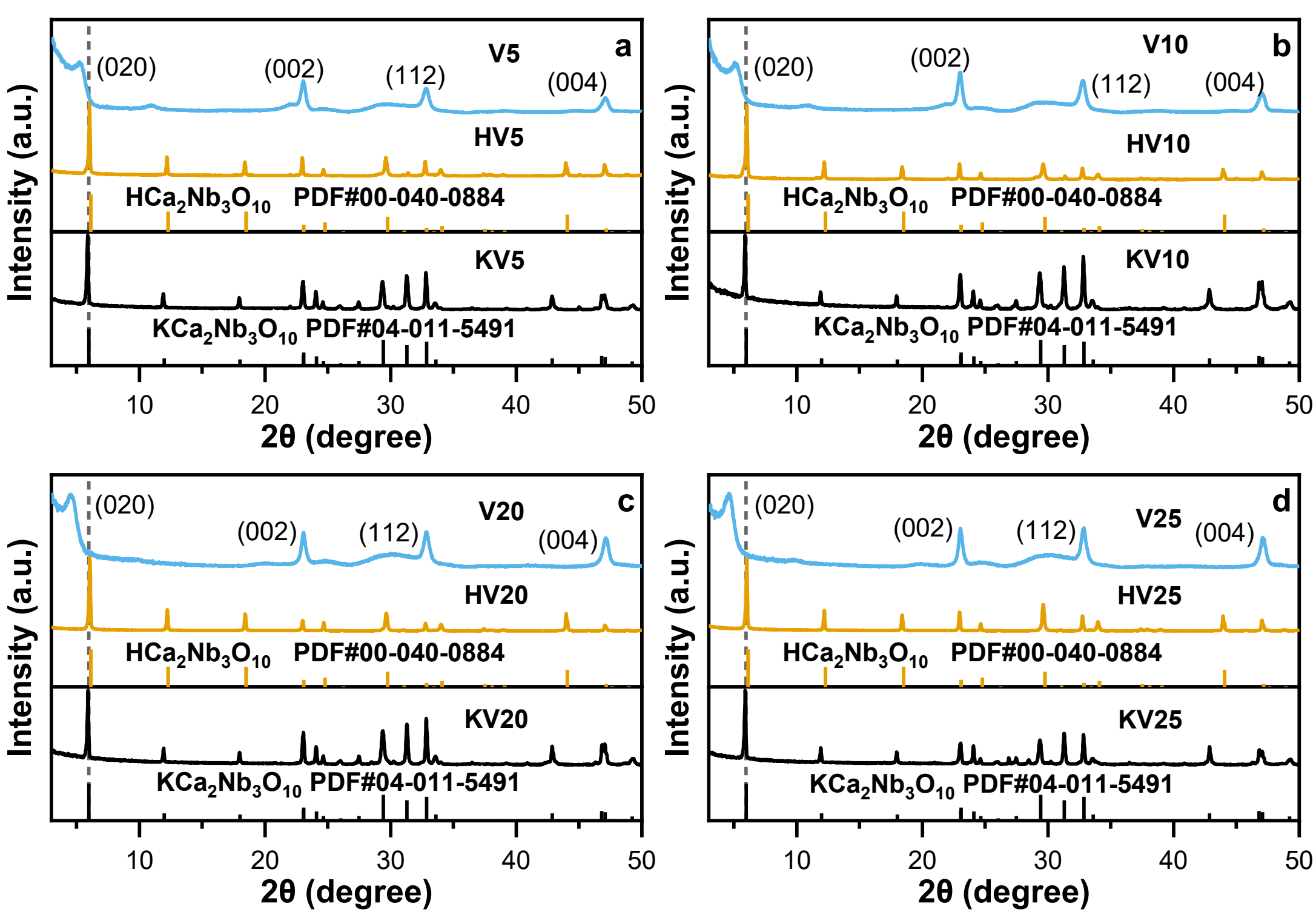

Figure S1. XRD diffractograms of parent, proton exchanged, and nanosheet forms of (a) x=0.15, (b) x=0.3, (c) x=0.6, (d) x=0.75 $KCa_2Nb_{3-x}V_xO_{10}$ compositions.

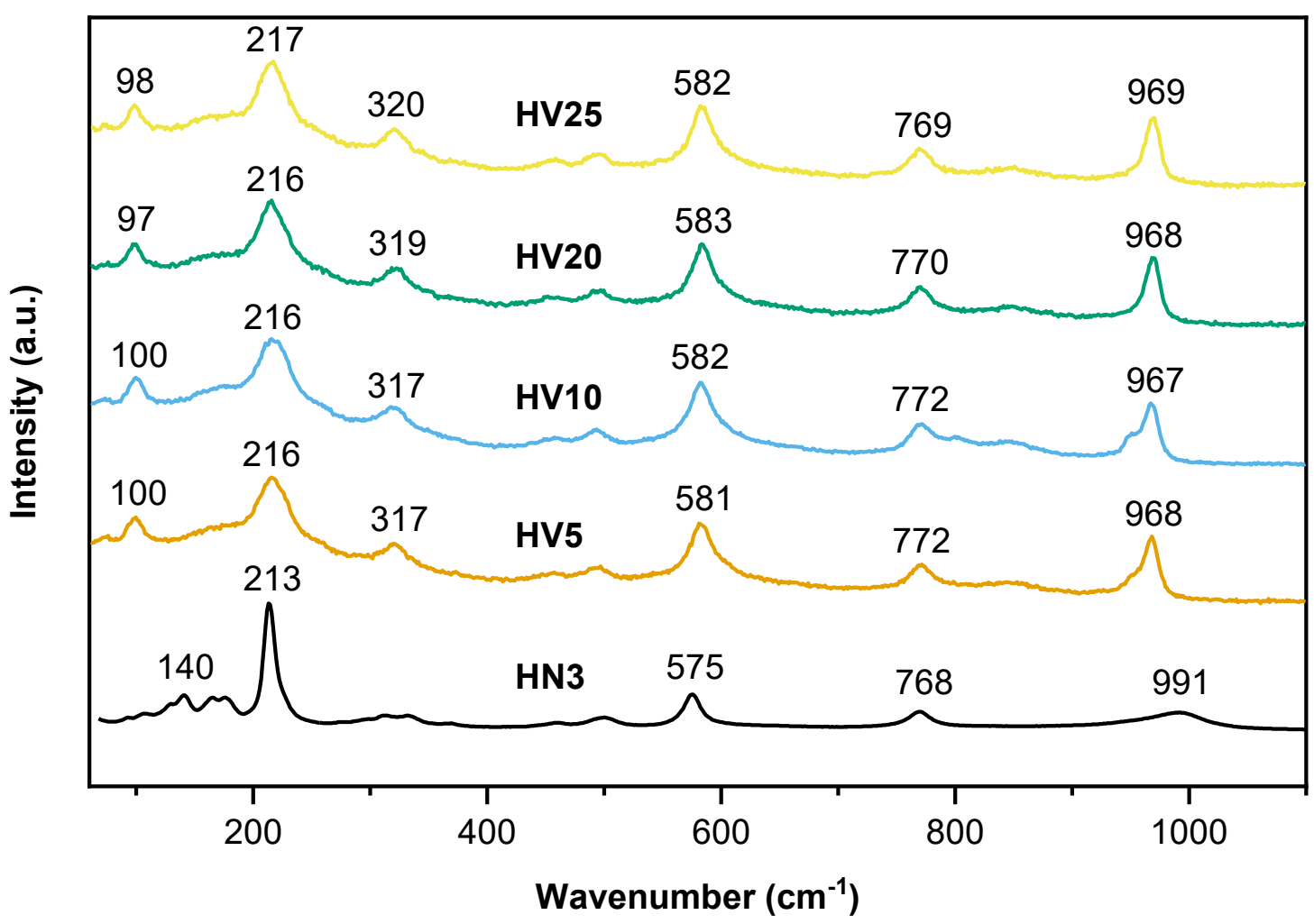

Figure S2. Raman spectra of $HCa_2Nb_{3-x}V_xO_{10}$ compositions.

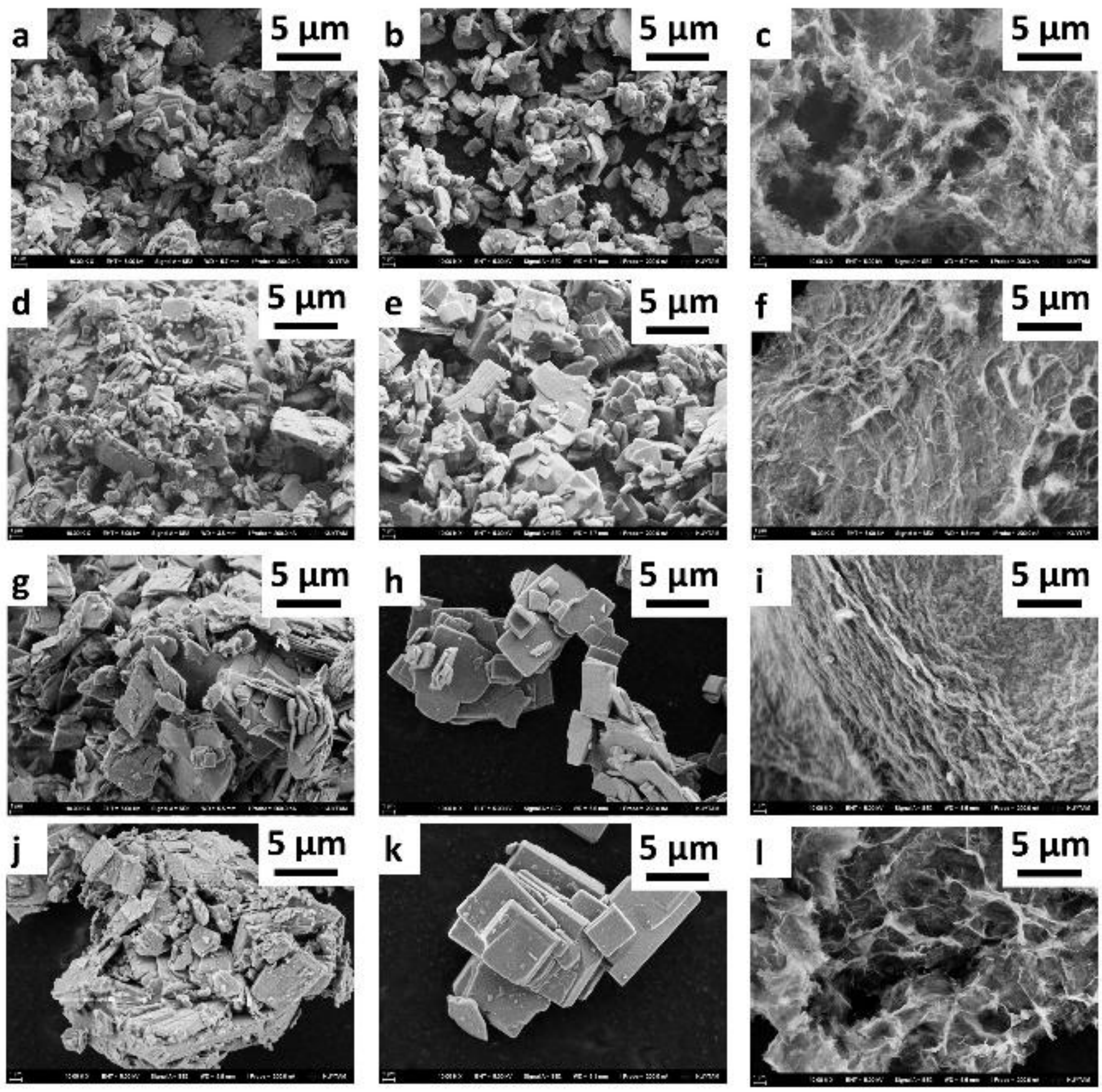

Figure S3. SEM images of (a) KN3, (b) HN3, (c) N3, (d) KV5, (e) HV5, (f) V5, (g) KV20, (h) HV20, (i) V20, (j) KV25, (k) HV25, (l) V25.

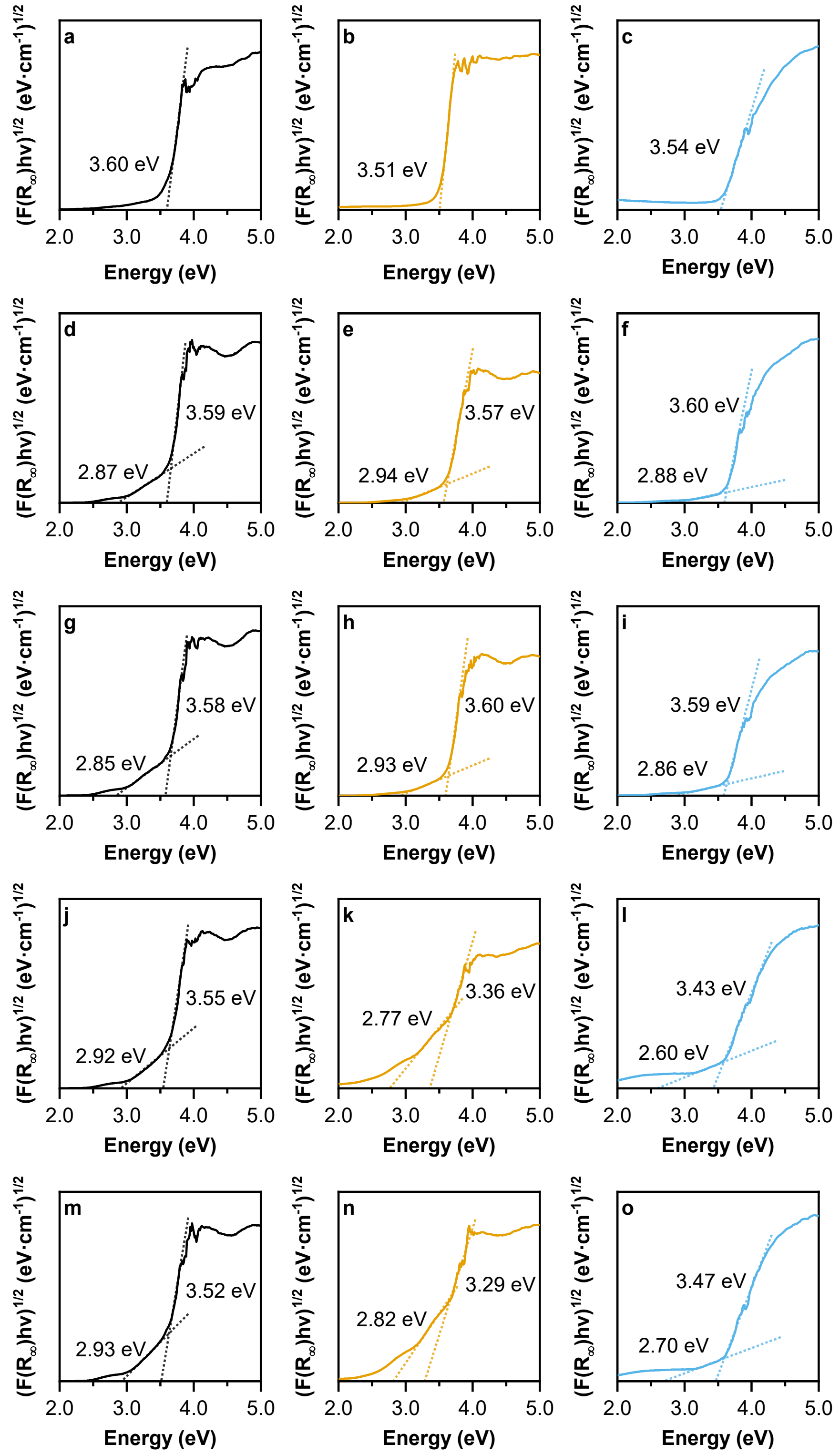

Figure S4. Tauc plots and linear fits of (a) KN3, (b) HN3, (c) N3, (d) KV5, (e) HV5, (f) V5, (g) KV10, (h) HV10, (i) V10, (j) KV20, (k) HV20, (l) V20, (m) KV25, (n) HV25, (o) V25.

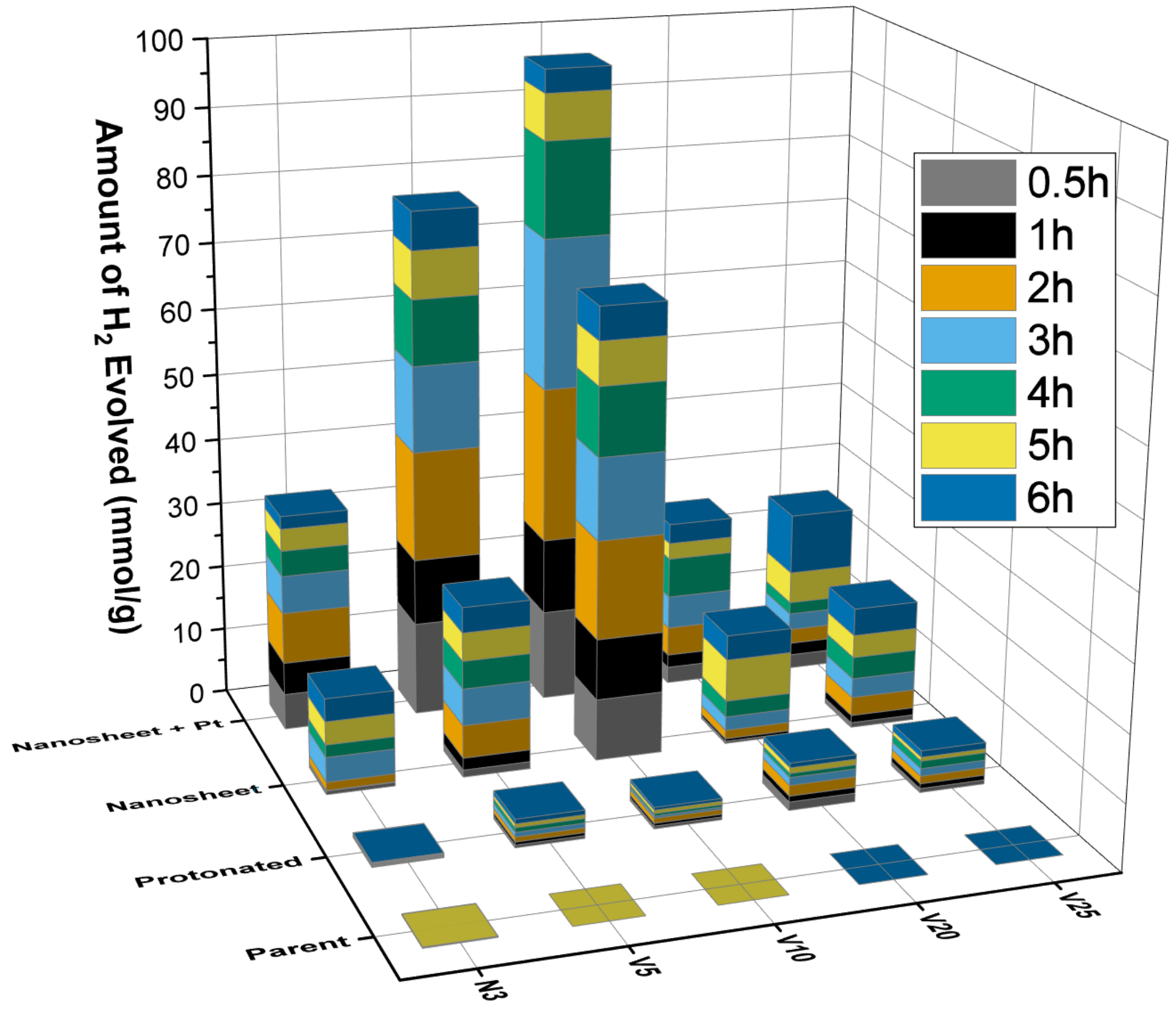

Figure S5. Hourly $H_2$ production amounts per gram of photocatalyst powder given for all phases and doping amounts.

**REFERENCES**

1. Fujishima, A. & Honda, K. Electrochemical Photolysis of Water at a Semiconductor Electrode. *Nature* https://doi.org/10.1038/238037a0 (1972) doi:10.1038/238037a0.
2. Kung, H. H., Jarrett, H. S., Sleight, A. W. & Ferretti, A. Semiconducting oxide anodes in photoassisted electrolysis of water. *J. Appl. Phys.* **48**, 2463–2469 (1977).
3. Mavroides, J. G., Kafalas, J. A. & Kolesar, D. F. Photoelectrolysis of water in cells with SrTiO3 anodes. *Applied Physics Letters* **28**, 241–243 (1976).
4. Watanabe, T., Fujishima, A. & Honda, K. Photoelectrochemical Reactions at SrTiO3 Single Crystal Electrode. *bull. Chem. Soc. Jpn.* **49**, 355–358 (1976).
5. Wrighton, M. S. *et al.* Strontium titanate photoelectrodes. Efficient photoassisted electrolysis of water at zero applied potential. *J. Am. Chem. Soc.* **98**, 2774–2779 (1976).
6. Wrighton, M. S., Wolczanski, P. T. & Ellis, A. B. Photoelectrolysis of water by irradiation of platinized n-type semiconducting metal oxides. *Journal of Solid State Chemistry* https://doi.org/10.1016/0022-4596(77)90185-2 (1977) doi:10.1016/0022-4596(77)90185-2.
7. Ellis, A. B., Kaiser, S. W. & Wrighton, M. S. Semiconducting potassium tantalate electrodes. Photoassistance agents for the efficient electrolysis of water. *J. Phys. Chem.* **80**, 1325–1328 (1976).
8. Butler, M. A. Photoelectrolysis and physical properties of the semiconducting electrode WO2. *J. Appl. Phys.* **48**, 1914–1920 (1977).
9. Hardee, K. L. & Bard, A. J. Semiconductor Electrodes: X . Photoelectrochemical Behavior of Several Polycrystalline Metal Oxide Electrodes in Aqueous Solutions. *J. Electrochem. Soc.* **124**, 215 (1977).
10. Kim, H. & Laitinen, H. A. Photoeffects at Polycrystalline Tin Oxide Electrodes. *J. Electrochem. Soc.* **122**, 53 (1975).
11. Scaife, D. E. Oxide semiconductors in photoelectrochemical conversion of solar energy. *Solar Energy* **25**, 41–54 (1980).
12. Wagner, F. T. & Somorjai, G. A. Photocatalytic and photoelectrochemical hydrogen production on strontium titanate single crystals. *J. Am. Chem. Soc.* **102**, 5494–5502 (1980).
13. Domen, K., Kudo, A., Onishi, T., Kosugi, N. & Kuroda, H. Photocatalytic decomposition of water into hydrogen and oxygen over nickel(II) oxide-strontium titanate (SrTiO3) powder. 1. Structure of the catalysts. *J. Phys. Chem.* **90**, 292–295 (1986).
14. Domen, K. *et al.* Photodecomposition of water and hydrogen evolution from aqueous methanol solution over novel niobate photocatalysts. *J. Chem. Soc., Chem. Commun.* 356–357 (1986) doi:10.1039/C39860000356.

15. Domen, K., Kudo, A., Tanaka, A. & Onishi, T. Overall photodecomposition of water on a layered niobiate catalyst. *Catalysis Today* **8**, 77–84 (1990).
16. Domen, K., Yoshimura, J., Sekine, T., Tanaka, A. & Onishi, T. A novel series of photocatalysts with an ion-exchangeable layered structure of niobate. *Catal Lett* **4**, 339–343 (1990).
17. Kato, H. & Kudo, A. New tantalate photocatalysts for water decomposition into H2 and O2. *Chemical Physics Letters* **295**, 487–492 (1998).
18. Ruddlesden, S. N. & Popper, P. New compounds of the $K_2NIF_4$ type. *Acta Cryst* **10**, 538–539 (1957).
19. Dion, M., Ganne, M. & Tournoux, M. Nouvelles familles de phases MIMII2Nb3O10 a feuillets "perovskites". *Materials Research Bulletin* **16**, 1429–1435 (1981).
20. Gopalakrishnan, J. & Bhat, V. A2Ln2Ti3O10 (A = potassium or rubidium; Ln = lanthanum or rare earth): a new series of layered perovskites exhibiting ion exchange. *Inorg. Chem.* **26**, 4299–4301 (1987).
21. Gopalakrishnan, J., Uma, S. & Bhat, V. Synthesis of Layered Perovskite Oxides, ACa2-xLaxNb3-xTixO10 (A = K, Rb, Cs), and Characterization of New Solid Acids, HCa2-xLaxNb3-xTixO10 (0 < x ≤ 2), Exhibiting Variable Bronsted Acidity. *ACS Publications* https://pubs.acs.org/doi/abs/10.1021/cm00025a025 (1993) doi:10.1021/cm00025a025.
22. Jacobson, A. J., Lewandowski, J. T. & Johnson, J. W. Ion exchange reactions of the layered solid acid HCa2Nb3O10 with alkali metal cations. *Materials Research Bulletin* **25**, 679–686 (1990).
23. Toda, K., Teranishi, T., Ye, Z.-G., Sato, M. & Hinatsu, Y. Structural chemistry of new ion-exchangeable tantalates with layered perovskite structure: new dion–jacobson phase MCa2Ta3O10 (M = alkali metal) and ruddlesden–popper phase Na2Ca2Ta3O10. *Materials Research Bulletin* **34**, 971–982 (1999).
24. Takata, T. *et al.* Photocatalytic Decomposition of Water on Spontaneously Hydrated Layered Perovskites. *Chem. Mater.* **9**, 1063–1064 (1997).
25. Brito, A. S. *et al.* Theoretical-experimental evaluation of the photocatalytic activity of KCa2Ta3−xNbxO10. *Materials Letters* **253**, 392–395 (2019).
26. Geselbracht, M. J. *et al.* New solid acids in the triple-layer Dion–Jacobson layered perovskite family. *Materials Research Bulletin* **46**, 398–406 (2011).
27. Han, Y.-S., Park, I. & Choy, J.-H. Exfoliation of layered perovskite, KCa2Nb3O10,into colloidal nanosheets by a novel chemical process. *J. Mater. Chem.* **11**, 1277–1282 (2001).
28. Ebina, Y., Sasaki, T., Harada, M. & Watanabe, M. Restacked Perovskite Nanosheets and Their Pt-Loaded Materials as Photocatalysts. *Chem. Mater.* **14**, 4390–4395 (2002).
29. Izawa, K. *et al.* Photoelectrochemical Oxidation of Methanol on Oxide Nanosheets. *J. Phys. Chem. B* **110**, 4645–4650 (2006).
30. Ida, S. *et al.* Preparation of a Blue Luminescent Nanosheet Derived from Layered Perovskite

Bi2SrTa2O9. *J. Am. Chem. Soc.* **129**, 8956–8957 (2007).

31. Ida, S. *et al.* Photoluminescence Spectral Change in Layered Titanate Oxide Intercalated with Hydrated Eu3+. *J. Phys. Chem. B* **110**, 23881–23887 (2006).
32. Singh, B. *et al.* Single-Atom (Iron-Based) Catalysts: Synthesis and Applications. *Chem. Rev.* **121**, 13620–13697 (2021).
33. Üstünel, T., Ide, Y., Kaya, S. & Doustkhah, E. Single-Atom Sn-Loaded Exfoliated Layered Titanate Revealing Enhanced Photocatalytic Activity in Hydrogen Generation. *ACS Sustainable Chem. Eng.* **11**, 3306–3315 (2023).
34. Maeda, K. & Domen, K. New Non-Oxide Photocatalysts Designed for Overall Water Splitting under Visible Light. *J. Phys. Chem. C* **111**, 7851–7861 (2007).
35. Maeda, K. & Domen, K. Photocatalytic Water Splitting: Recent Progress and Future Challenges. *J. Phys. Chem. Lett.* **1**, 2655–2661 (2010).
36. Ida, S., Okamoto, Y., Matsuka, M., Hagiwara, H. & Ishihara, T. Preparation of Tantalum-Based Oxynitride Nanosheets by Exfoliation of a Layered Oxynitride, CsCa2Ta3O10–xNy, and Their Photocatalytic Activity. *J. Am. Chem. Soc.* **134**, 15773–15782 (2012).
37. Ida, S., Okamoto, Y., Koga, S., Hagiwara, H. & Ishihara, T. Black-colored nitrogen-doped calcium niobium oxide nanosheets and their photocatalytic properties under visible light irradiation. *RSC Adv.* **3**, 11521–11524 (2013).
38. da Silva Maia, A. *et al.* Preparation of niobium based oxynitride nanosheets by exfoliation of Ruddlesden-Popper phase precursor. *Solid State Sciences* **54**, 17–21 (2016).
39. Tallal, M. *et al.* Review: Vanadium pentoxide based catalysts for photocatalytic water splitting. *J Mater Sci* **61**, 5008–5031 (2026).
40. Li, Y. *et al.* Research progress of vanadium pentoxide photocatalytic materials. *RSC Adv.* **13**, 22945–22957 (2023).
41. Haldar, K. K. *et al.* Efficient MoS2/V2O5 Electrocatalyst for Enhanced Oxygen and Hydrogen Evolution Reactions. *Electrocatalysis* **14**, 624–635 (2023).
42. Sarala, S., Karthik, P., Sasikala, V., Prakash, N. & Mukkannan, A. V2O5 -doped NiFe-layered double hydroxides: bifunctional catalysts for energy and environmental remediation. *Res Chem Intermed* **51**, 2955–2979 (2025).
43. Sasikumar, K. & Ju, H. Recent Advances in Vanadate-Based Materials for Photocatalytic Hydrogen Production. *Molecules* **30**, 789 (2025).
44. Antony, A. J., Jelastin Kala, S. M., Joel, C., Bennie, R. B. & Praveendaniel, S. Enhancing the visible light induced photocatalytic properties of WO3 nanoparticles by doping with vanadium. *Journal of Physics and Chemistry of Solids* **157**, 110169 (2021).
45. Wang, B., Zhang, G., Leng, X., Sun, Z. & Zheng, S. Characterization and improved solar light activity of

vanadium doped $TiO_2$/diatomite hybrid catalysts. *Journal of Hazardous Materials* **285**, 212–220 (2015).

46. Lin, W.-C. & Lin, Y.-J. Effect of Vanadium(IV)-Doping on the Visible Light-Induced Catalytic Activity of Titanium Dioxide Catalysts for Methylene Blue Degradation. *Environmental Engineering Science* **29**, 447–452 (2012).
47. El Sayed, A. M. & Alanazi, F. K. Microstructural characterization, electrical, and optical study of $V_2O_5$-doped $Cr_2O_3$ films for photonic applications. *J Mater Sci: Mater Electron* **35**, 2164 (2024).
48. Wang, R. *et al.* Cobalt-doped $V_2O_5$ hexagonal nanosheets for superior photocatalytic toxic pollutants degradation, Cr (VI) reduction, and photoelectrochemical water oxidation performance. *Environmental Research* **217**, 114923 (2023).
49. Bantawal, H., Shenoy, U. S. & Bhat, D. K. Vanadium-Doped $SrTiO_3$ Nanocubes: Insight into role of vanadium in improving the photocatalytic activity. *Applied Surface Science* **513**, 145858 (2020).
50. Fukuoka, H., Isami, T. & Yamanaka, S. Crystal Structure of a Layered Perovskite Niobate $KCa_2Nb_3O_{10}$. *Journal of Solid State Chemistry* **151**, 40–45 (2000).
51. Shelyapina, M. G. *et al.* 1H NMR Study of the $HCa_2Nb_3O_{10}$ Photocatalyst with Different Hydration Levels. *Molecules* **26**, 5943 (2021).
52. Chen, Y. *et al.* Structure and dehydration of layered perovskite niobate with bilayer hydrates prepared by exfoliation/self-assembly process. *Journal of Solid State Chemistry* **181**, 1684–1694 (2008).
53. Jehng, J. M. & Wachs, I. E. Structural chemistry and Raman spectra of niobium oxides. *Chem. Mater.* **3**, 100–107 (1991).